# Robust mode-locking in a hybrid ultrafast laser based on nonlinear multimodal interference

Xuanyi Liu[a], Maolin Dai[a], Denghui Pan[a], Kaibin Lin[b], Boris A. Malomed[c,d], Qian Li[b*], H. Y. Fu[a,**]

[a] *Tsinghua Shenzhen International Graduate School and Tsinghua-Berkeley Shenzhen Institute, Tsinghua University, Shenzhen 518055, China*
[b] *School of Electronic and Computer Engineering, Peking University, Shenzhen 518055, China*
[c] *Department of Physical Electronics, School of Electrical Engineering, Faculty of Engineering, and Center for Light-Matter Interaction, Tel Aviv University, Tel Aviv 69978, Israel*
[d] *Instituto de Alta Investigación, Universidad de Tarapacá, Casilla 7D, Arica, Chile*
\* Corresponding author: liqian@pkusz.edu.cn
\*\*Corresponding author: hyfu@sz.tsinghua.edu.cn

ABSTRACT

We experimentally demonstrate the realization of a half-polarization-maintaining (half-PM) fiber laser, in which mode-locking is provided by a reflective multimode-interference saturable absorber (SA). In the specially designed SA, linearly polarized light is coupled into a 15-cm-long graded-index multimode fiber (GIMF) through the PM fiber, and then reflected back to the PM structure through a mirror pigtailed with a single-mode fiber (SMF). The modulation depth and saturation peak power are measured to be 1.5% and 0.6 W, respectively. The proposed SA device is incorporated into a novel half-PM erbium-doped fiber oscillator, which generates soliton pulses with 409 fs temporal duration at a 33.3 MHz repetition rate. The proposed fiber laser is compared with a conventional non-PM fiber laser mode-locked by nonlinear polarization evolution (NPE) in terms of optical properties such as spectral bandwidth, pulse duration, and stability performance. Short- and long-time stability tests and superior noise performance corroborate robust mode-locking in this setup.

## 1. Introduction

Advanced mode-locked fiber oscillators with enhanced performance and stability are considered as valuable assets for numerous applications, including terahertz-wave generation [1], laser micromachining [2], and medical treatment [3] due to their structural compactness, low fabrication cost, and superior thermo-optical characteristics [4]. The growing stringent demand for industrial and scientific applications is currently driving the research into the field of ultrashort pulse generation in laser cavities. The improvement of output pulse performance and enhancement of environmental stability are crucial aspects of the application of fiber lasers in complex environments. Various pulse picking methods, which have been developed to achieve stable mode-locking, critically depend on employed saturable absorbers (SAs). Among them, two-dimensional materials [5] and the semiconductor saturable absorber mirror (SESAM) [6] enable self-starting operation but suffer from low damage threshold and performance degradation over long-term operation. Additive pulse mode-locking (APM) techniques, including nonlinear polarization evolution (NPE) [7], nonlinear optical loop mirror (NOLM) [8], and nonlinear amplifying loop mirror (NALM) [9], are utilized as effective SAs in mode-locked fiber lasers. However, the precise control of the polarization states in NPE fiber lasers and extremely challenging self-starting issues in figure-8 fiber lasers hinder their widespread use.

Recently, much interest has been drawn to designing and constructing mode-locked lasers incorporating a segment of graded-index multimode fiber (GIMF) as an all-fiber SA based on nonlinear multimodal interference (NL-MMI) technique, which enables many important advances [10,11,20–24,12–19]. GIMF-based SA was initially proposed by Nazemosadat et al. [10] and experimentally elaborated in Yb-doped [11], Er-doped [12,13,17,18], Tm-doped [14,15], and dual-color [16] mode-locked fiber lasers. By stretching the GIMF, Chen et al. have investigated the coexistence of conventional high-energy solitons and stretched pulses in an Er-doped mode-locked fiber laser [17]. Further, it was demonstrated that the GIMF-based SA features high-power tolerance and large modulation depth, thus high-energy soliton pulses are generated in a dual-pumped GIMF-based fiber laser [13]. Another major advantage of such a light source is that it can achieve tunable mode-locking wavelength by adjusting the intra-cavity loss, which is beneficial to realizing multi-wavelength mode-locking [18]. In addition, the GIMF-based SA can function as a bandpass filter, which is necessary for the generation of dissipative solitons [11]. However, previously reported NL-MMI-based fiber lasers have been implemented in non-polarization-maintaining (non-PM) fibers, making the mode-locking states highly sensitive to environmental perturbations. The use of PM fiber may help to solve this problem. Nevertheless, a difficulty is that polarization controllers (PCs) are critically important for tuning intra-cavity birefringence and optimizing mode-locking, which contradicts the use of the PM fibers. Thus, exploring this mode-locking mechanism in combination with PM fibers is of great interest for the design and applications of NL-MMI-based fiber lasers. Adopting long PM fibers in the laser cavities can reduce the laser's susceptibility to environmental factors, such as temperature changes and mechanical vibrations. Previously, the NPE mode-locking mechanism was demonstrated to be feasible in half-PM laser cavities [25,26]. Another impressive work is that chirp-free solitons are directly generated in a normal-dispersion fiber laser containing a few meters of PM fiber. Fiber birefringence induced by the PM fiber, normal-dispersion, and nonlinear effect follow a phase-matching principle, enabling the formation of the near-chirp-free soliton [27]. Whether developing new pulse formation mechanisms in fiber lasers or improving the environmental stability of fiber lasers, the use of PM fibers can be of great benefit. Up to now, there have been no reports on employing the NL-MMI technique in the hybrid-structure fiber lasers to realize ultrashort pulse generation.

This work reports the creation and operation of a simple and compact Er-doped fiber laser with a half-PM and half-ring structure, which makes use of a reflective-type NL-MMI saturable absorber. The laser emits ultrashort pulses with the 409 fs temporal width at a 33.3 MHz repetition rate. The use of PM fibers greatly enhances the laser's practicability -- in particular, its stability against environmental perturbations. Benefiting from the robust configuration, low output power fluctuations of 0.62% the relative root mean square (RMS) in the course of 24 hours are secured. Moreover, 1.045 ps timing jitter and 0.057% relative intensity noise (RIN) integrated from 1 kHz to 10 MHz are demonstrated. In contrast, the optical performance of the proposed half-PM GIMF-based fiber laser is superior to that of a non-PM NPE-based fiber laser. To the best of our knowledge, this is the first laser configuration for implementing the NL-MMI mode-locking mechanism in PM fibers. Such an environmentally stable fiber laser offers the potential for use as a pulsed laser source in ultrafast technologies.

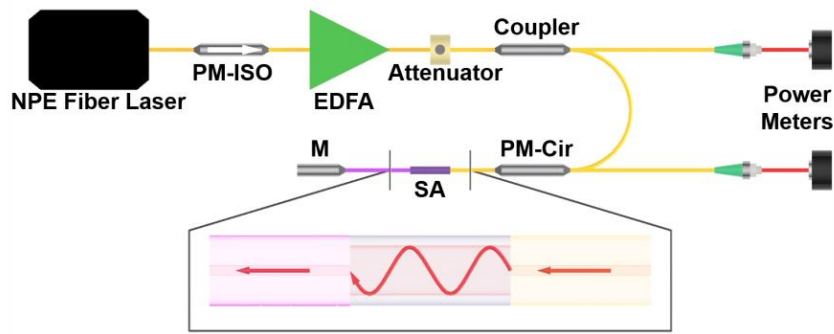

**Fig. 1.** The measurement of saturable absorption using the balanced twin-detector system. NPE: nonlinear polarization evolution; ISO: isolator; EDFA: erbium-doped fiber amplifier; Cir: circulator; M: mirror.

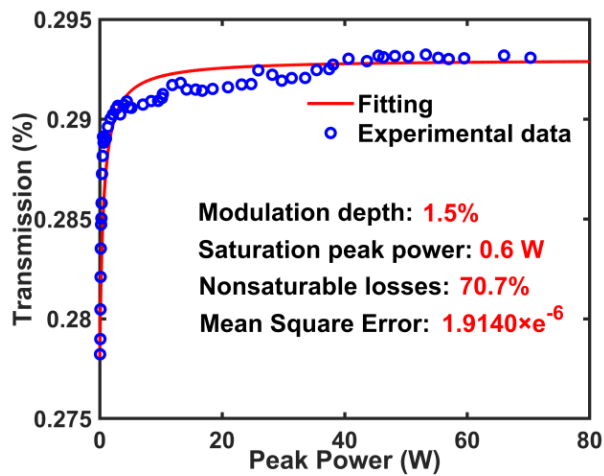

**Fig. 2.** The nonlinear saturable absorption curve of the reflective-type NL-MMI saturable absorber.

## 2. Characterization of the GIMF SA

To characterize the performance of the SA, a balanced twin-detector system with two paths and two power meters is constructed, as shown in Fig. 1. The SA is fabricated as a short segment of commercially available GIMF with a length of 15 cm and a core diameter of 62.5 μm. Its ends are spliced into a single-mode fiber (SMF), and into the PM one. Seed pulses with a repetition rate of 117 MHz are produced by a homemade PM-NPE fiber laser [28], and then amplified by the PM erbium-doped fiber amplifier (EDFA) to attain sufficiently high pulse energy. Following the amplification, the optical signal with 1.05-ps temporal duration is attenuated and split into two beams by a 10:90 coupler. Amplified pulses carrying 90% energy are injected into the SA through a PM circulator (PM-Cir), and finally reflected by the mirror (M) back to the power meter for the detection. The fast axis of the PM-Cir is blocked, ensuring that only linearly polarized light passes through the SA. Except for the pigtail of the mirror, which is based on the SMF, pigtails of other optical elements use PM fibers, which is helpful for the stability of the test, which verifies that the proposed SA can be used in the half-PM laser structure. Figure 2 displays the measured nonlinear saturable absorption data, fitted by the following equation:

$$T(P_{peak}) = 1 - \frac{\Delta T}{1 + P_{peak}/P_{sat}} - \alpha_{ns}, \quad (1)$$

where $T(P_{peak})$ represents the transmission, $P_{peak}$ is the peak power of light, $P_{sat}$ is the saturation peak power, $\Delta T$ is the modulation depth, and $\alpha_{ns}$ stands for nonsaturable losses. From the fitting results, it can be seen that the proposed SA features a modulation depth of 1.5%. The saturation peak power and nonsaturable losses are 0.6 W and 70.7%, respectively. The Mean Square Error (MSE) to estimate the error between the fitted and original data is computed to be $1.9140 \times e^{-6}$.

## 3. Laser structures and lasing performance

These remarkable nonlinear properties indicate that the NL-MMI-based SA has the potential for pulse-picking in mode-locked fiber lasers. The experimental schematic of the half-PM fiber laser, which incorporates the NL-MMI-based SA, is displayed in Fig. 3. The laser cavity is composed of dominant PM devices and a few non-PM ones. A segment of the PM erbium-doped fiber (PM-ESF-7/125, Nufern) is employed as the gain medium with the length of 72 cm, which has an absorption coefficient of 24 dB/m at 976 nm, being core-pumped by a laser diode (LD) at this wavelength through a PM wavelength division multiplexer (WDM). A 1550 nm PM circulator (Cir) performs dual functions in the laser cavity. First, it acts as an isolator to ensure the counterclockwise propagation of pulses in the ring, as indicated by the red arrow in Fig. 3. The second function of the PM circulator is to realize the combination of the reflective-type SA with the half-PM laser cavity. The 10:90 output coupler (OC) inside the ring is utilized to extract 10% light from the laser cavity. Fast axes with all PM fiber pigtailed devices are blocked and ultrashort pulses propagate only along the slow axes of PM fibers. When the pulses enter the PM portion of the fiber laser, the PM-WDM with only slow-axis working serves as a polarizer, ensuring a stable generation of linearly polarized pulses in the fiber oscillator. As for the non-PM portion, a polarization controller (PC) is exploited to tune the light polarization. The end mirror (M) reflects the pulse and thus closes a loop. To enhance the environmental stability of the fiber laser, the length of the non-PM fiber is made as short as possible. The lengths of the passive PM and non-PM fibers are 135 cm and 127 cm, respectively. The group velocity dispersions (GVD) of the passive PM fiber, active one, and SMF are -22.9 ps$^2$/km, -20.4 ps$^2$/km, and -22.9 ps$^2$/km, respectively, leading to an estimated intra-cavity net dispersion of -0.074 ps$^2$. Thus, the laser operates in the anomalous-GVD regime, supporting the conventional soliton generation.

To highlight the superior performance of the designed half-PM GIMF fiber laser, a conventional NPE fiber laser composed entirely of non-PM fiber devices is constructed for comparative analysis, as shown in Fig. 4. The non-PM NPE fiber laser also works in the anomalous dispersion regime. By precisely controlling the length of the laser cavity, ultrashort pulses produced by the non-PM NPE laser outputs can have the same repetition rate as the half-PM GIMF laser, and consequently, the average output power and single-pulse energy of both lasers are in the same order of magnitude. This ensures a reasonable comparison of lasing performance.

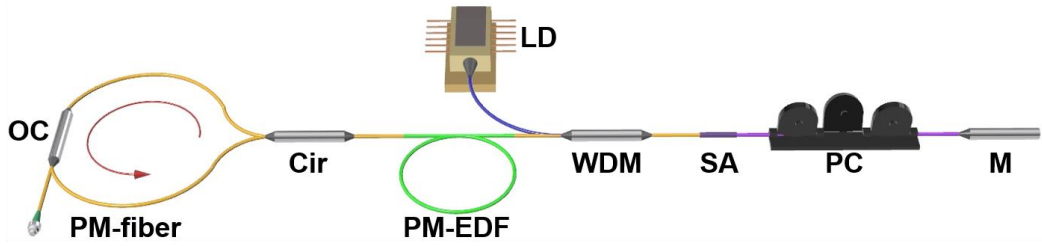

**Fig. 3.** The schematic of the half-PM fiber laser with NL-MMI technique. LD: laser diode; OC: output coupler; Cir: circulator; PM-EDF: polarization-maintaining erbium-doped fiber; WDM: wavelength-division multiplexer; SA: saturable absorber; PC: polarization controller; M: mirror.

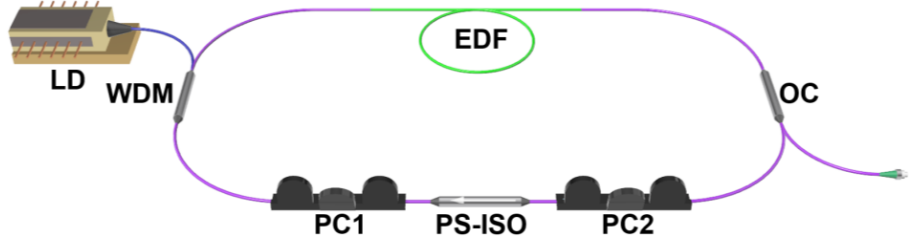

**Fig. 4.** The schematic of the non-PM fiber laser with NPE technique. LD: laser diode; WDM: wavelength-division multiplexer; EDF: erbium-doped fiber; OC: output coupler; PC: polarization controller; PS-ISO: polarization-sensitive isolator.

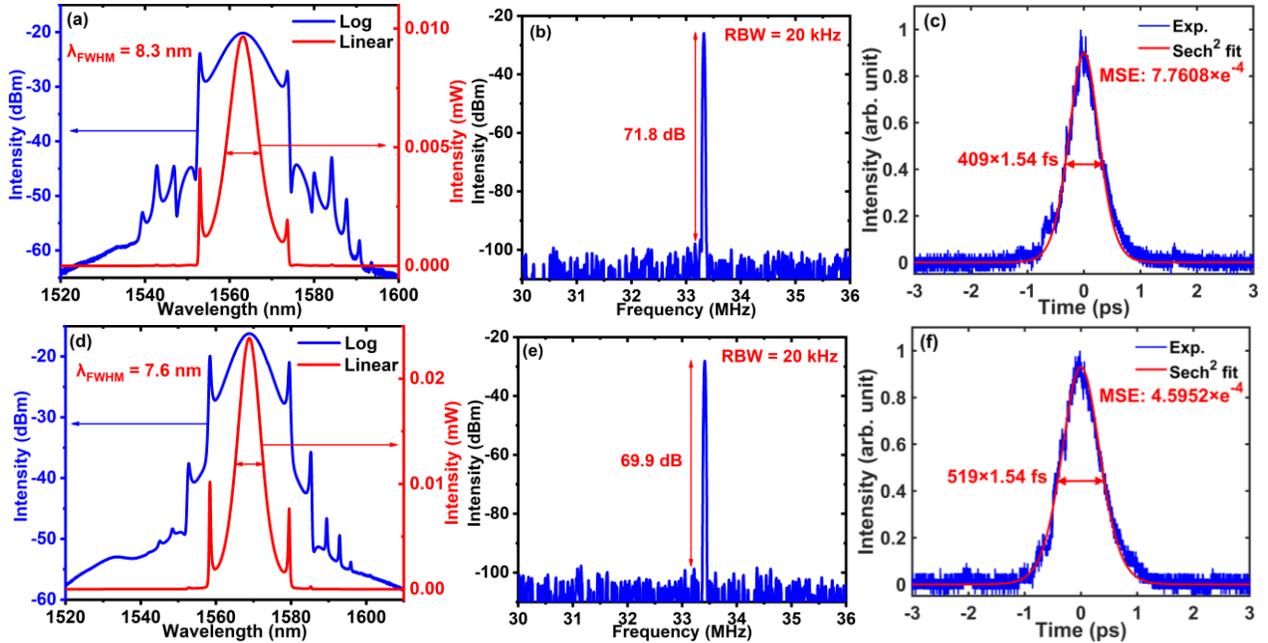

**Fig. 5.** (a)-(c) Single-pulse emission from the half-PM GIMF fiber laser with a pump power of 68 mW; (d)-(f) Single-pulse emission from the non-PM NPE fiber laser with a pump power of 75 mW; (a), (d) The optical spectrum, displayed on the logarithmic and linear scales; (b), (e) The radio frequency (RF) spectrum; (c), (f) The autocorrelation trace.

When paddles of the PC are adjusted appropriately, self-starting mode-locking can be initiated but requires a slightly higher pump power exceeding 180 mW. Decreasing the power to 68 mW enables clean single-pulse operation. The primary mode-locking performance of the generated single pulse is summarized in Fig. 5. Figure 5(a) shows a typical broadband optical spectrum with obvious Kelly-sidebands measured by an optical spectrum analyzer (OSA, AQ6370D, Yokogawa). The full-width half-maximum (FWHM) of the spectrum, defined as the peak width at half the height of the linear spectral peak, is measured to be 8.3 nm with a central wavelength of 1563.1 nm. A radio frequency (RF) signal analyzer (N9030B, Agilent) with a bandwidth from 3 to 50 GHz is employed to measure the RF spectrum, exhibiting a peak at the fundamental repetition rate of 33.3 MHz, which agrees well with the roundtrip time (30 ns) of the ultrashort pulses in the laser cavity. A high signal-to-noise ratio (SNR) of 71.8 dB has been demonstrated at a resolution bandwidth (RBW) of 20 kHz, verifying the stable passive mode-locking. The corresponding autocorrelation trace of the delivered ultrashort pulse is recorded by a commercial autocorrelator (FR-103XL/IR/FA) and displayed in Fig. 5(c). The pulse duration is estimated to be 409 fs, assuming a sech$^2$ pulse profile. The time-bandwidth product (TBP) is calculated to be 0.417, which is slightly larger than the Fourier-transform limit. The dispersion of the connection fiber used for the measurement is mainly responsible for the large pulse width. The formed solitons are slightly chirped due to the net anomalous dispersion of the laser cavity.

The lasing performance of the non-PM NPE fiber laser is presented in Figs. 5(d)-(f). At a pump power of 75 mW, the generated pulses characterize a spectral FWHM of 7.6 nm, which is narrower than that of Fig. 5(a). Correspondingly, a pulse duration of 519 fs is realized according to the autocorrelation trace shown in Fig. 5(f). Figure 5(e) depicts the RF spectrum with a lower SNR of 69.9 dB, thus demonstrating the more stable pulsed operation of the half-PM GIMF fiber laser.

The stability and reliability of the present half-PM fiber laser are characterized by long-term and short-term spectral properties, fluctuations of the average output power, and noise performance. All tests were conducted at room temperature with the fiber laser exposed to open air, without any sheltering box. The optical spectrum of the generated pulses is monitored over a long period of time over 2 hours. The shape, intensity, and bandwidth of the spectrum remain unchanged, demonstrating the long-term operation capability of the proposed laser. In particular, shaking the PM fiber does not affect the mode-locking state, a feature which was not demonstrated by previously reported non-PM GIMF-based fiber lasers [11–18]. The short-term stability test in the frequency domain was performed by means of the recently developed dispersive Fourier transformation (DFT) technique [29]. The optical spectrum is mapped onto the temporal waveform by time-stretching the ultrashort pulses using a dispersive element. Specifically, in our DFT setup, it is a 3-km-long dispersion compensation fiber (DCF, -160 ps/nm/km), which provides a total GVD of -480 ps/nm. The real-time spectra are detected by a fast photodetector (PD, UPD-15-IR2-FC, Alphalas) and then captured by a high-speed oscilloscope (MSO73304DX, Tektronix) with a bandwidth of 33 GHz. Figure 6(a) illustrates 5000 successive roundtrips of the DFT signal. The white curve in Fig. 6(a) indicates the corresponding energy evolution produced by integrating the real-time spectra. The virtually unchanged spectrum and smooth energy curve corroborate the remarkably high stability of the mode-locked pulses produced by the half-PM laser cavity. The time-averaged optical spectrum (red curve) measured by the OSA shows excellent agreement with the single-shot spectrum (the black curve) recorded by the DFT setup, as shown in the right inset of Fig. 6(a). Moreover, the average output power stability over 24 hours was measured using a power meter (S148 C, Thorlabs), shown in Fig. 6(b). The residual level of the instability is secured to be as low as 0.003 mW, with the average output power of 0.482 mW, and the corresponding RMS is 0.62%. For the non-PM NPE fiber laser, the power instability of 0.013 mW represents a relative RMS of 2.03% at an average output power of 0.641 mW. Thus, the fluctuations of the average output power are effectively suppressed in the half-PM laser configuration. The lasing performance comparison between the half-PM GIMF fiber laser and the non-PM NPE fiber laser is shown in Table 1.

The noise performance should be evaluated to facilitate high-precision applications of the fiber laser. In the free-running regime without any active noise suppression, the phase and amplitude noise were characterized using a phase noise analyzer (FSWP8, Rohde & Schwarz). The so-measured phase noise and its integration (the timing jitter) are reported in Fig. 7(a). The phase noise spectrum decreases from −88 dBc/Hz to −156 dBc/Hz for the offset frequency varying from 10 Hz to 10 MHz. It is inferred that the low-frequency noise is the main contributor to the timing jitter. Figure 7(b) shows the power spectral density of RIN and integrated RIN. Two main drops can be observed in the presented frequency range. The amplitude-noise curve slowly decreases from -113 dBc/Hz to -127 dBc/Hz in the interval of the offset frequency between 10 Hz and 300 kHz, followed by a rapid drop to -152 dBc/Hz as the offset frequency grows from 300 kHz to 10 MHz. Besides, multiple noise spikes in the frequency range from 10 Hz to 1 kHz are attributed to acoustic noise or mechanical vibrations [30]. The timing jitter and integrated RIN are 1.045 ps and 0.057%, respectively, when integrated with the range of the offset frequency from 1 kHz to 10 MHz. The noise performance may be further improved by shortening the cavity length to increase the repetition rate or performing delicate dispersion management [31]. The noise spectra and their corresponding integrations of the ultrashort pulses delivered by the non-PM NPE fiber laser are illustrated in Figs. 7(c)-(d). Table 1 also compares the timing jitter and integrated RIN over different offset frequency ranges. The phase noise of the non-PM NPE fiber laser in the low offset frequency range is much higher than that of the half-PM GIMF fiber laser, while the noise curves are at the same level in the high offset frequency range. The obtained results demonstrate that the use of PM fibers effectively reduces the disturbance to the fiber laser caused by mechanical vibration, temperature changes and other environmental factors, thereby suppressing the noise at low offset frequencies.

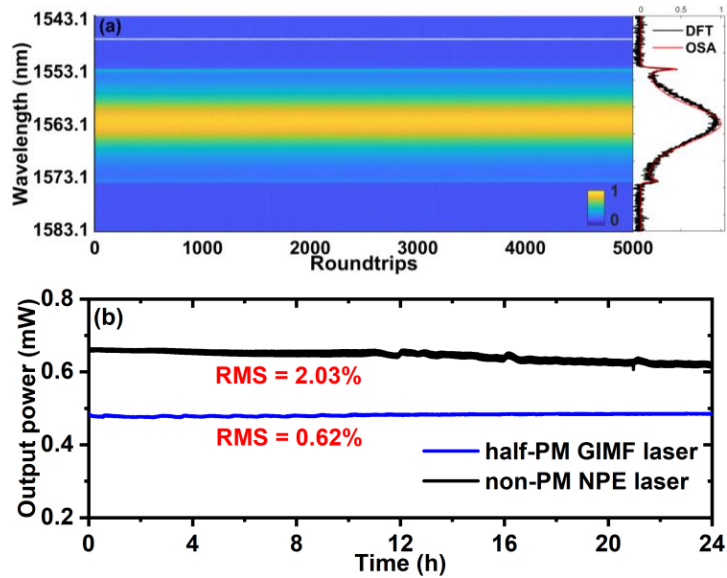

**Fig. 6.** (a) The short-term spectral stability tested by dint of the using DFT technique; (b) Output power fluctuations of in the course of 24 hours.

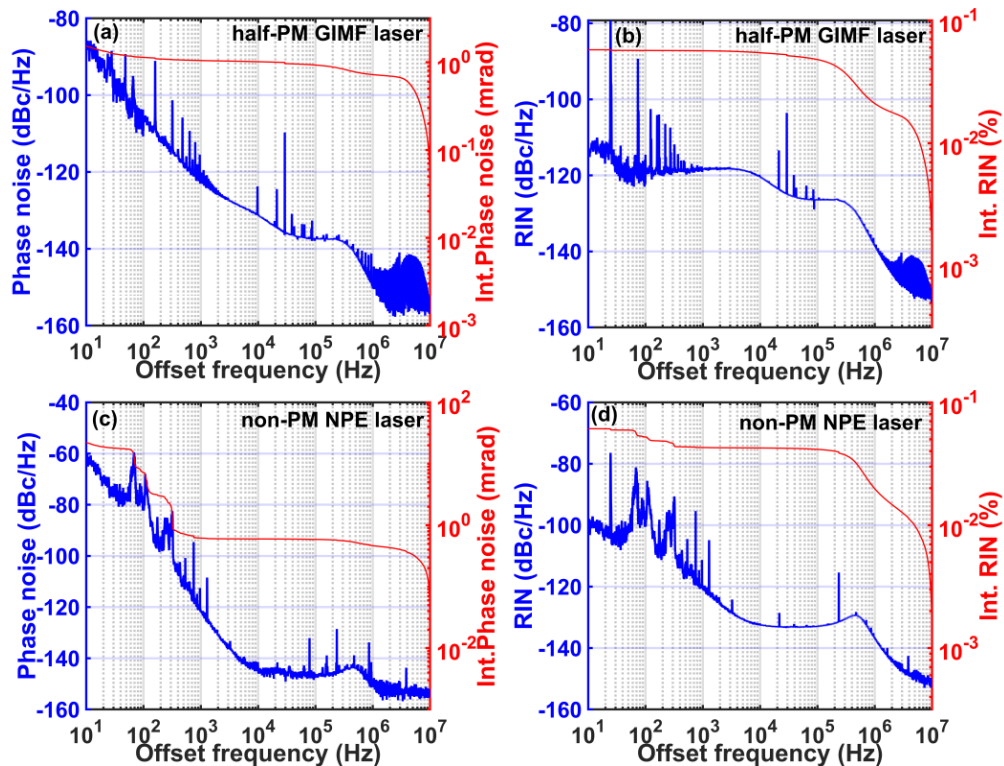

**Fig. 7.** Noise performance comparison between the proposed half-PM GIMF laser and a non-PM NPE laser. (a)-(b) The power spectral density of the phase noise, RIN, and their integrations of the proposed half-PM GIMF laser; (c)-(d) The power spectral density of the phase noise, RIN, and their integrations of a non-PM NPE laser.

**Table 1** Pulse performance comparison between the half-PM GIMF laser and the non-PM NPE laser

| | $\lambda_{FWHM}$ (nm) | RF spectra SNR (dB) | $\tau_{FWHM}$ (fs) | Average output power RMS (24 h) | Timing jitter (fs) [10 Hz-10 MHz] | Timing jitter (fs) [1 kHz-10 MHz] | Integrated RIN [10 Hz-10 MHz] | Integrated RIN [1 kHz-10 MHz] |
|---|---|---|---|---|---|---|---|---|
| Half-PM GIMF laser | 8.3 | 71.8 | 409 | 0.62% | 1.528 | 1.045 | 0.058% | 0.057% |
| Non-PM NPE laser | 7.6 | 69.9 | 519 | 2.03% | 22.381 | 0.617 | 0.061% | 0.043% |

$\lambda_{FWHM}$: spectral full-width half-maximum; $\tau_{FWHM}$: temporal full-width half-maximum.

## 4. Conclusion

In conclusion, we have built the first half-PM fiber oscillator based on the NL-MMI technique. The designed fiber laser produces 409-fs, 14-pJ conventional soliton pulses, whose fundamental repetition rate and center wavelength are 33.3 MHz and 1563.1 nm, respectively. The utilization of the PM fiber provides the enhancement of the environmental stability of the laser. Compared to other non-PM laser cavities such as conventional NPE fiber lasers, it exhibits excellent performance with a tiny output power fluctuation at the level of 0.62%. Low timing jitter of 1.045 ps and integrated RIN of 0.057% have been realized in the range of 1 kHz to 10 MHz. The presented half-PM fiber laser offers a stable pulsed light source for optical metrology and other applications. The approach can be extended to fabricate NL-MMI-based fiber lasers in other wavelength bands, high-energy fiber lasers, and multi-wavelength fiber lasers with robust mode-locking and enhanced stability.

## Funding

This work was supported by Overseas Research Cooperation Fund of Tsinghua Shenzhen International Graduate School (HW2020006), Shenzhen Fundamental Research Program (GXWD20201231165807007-20200827130534001), Youth Science and Technology Innovation Talent of Guangdong Province (2019TQ05X227), Israel Science Foundation (1286/17).

## Declaration of competing interest

The authors declare no conflicts of interest.